# Almeida-Thouless Line in BiFeO$_3$: Is Bismuth Ferrite a mean field spin glass?

**Manoj K Singh[1], Ram S. Katiyar[1], W. Prellier[2], and J F Scott[3]**

[1]Department of Physics and Institute of Functional Nano Materials, University of Puerto Rico, PR 00931-3343, USA
[2]Laboratoire CRISMAT, CNRS UMR 6508, ENSICAEN, 6 Bd du Maréchal Juin, F-14050 Caen Cedex, France
[3]Department of Earth science, University of Cambridge, Cambridge CB2, 3EQ,UK

E-mail: jsco99@esc.cam.ac.uk and rkatiyar@speclab.upr.edu

**Abstract**
Low-temperature magnetic properties of epitaxial BiFeO$_3$ (BFO) thin films grown on (111) SrTiO$_3$ substrates have been studied. Zero-field-cooled (ZFC) and field-cooled (FC) magnetization curves show a large discrepancy beginning at a characteristic temperature $T_f$ that is dependent on the magnetic-field strength. $T_f(H)$ varies according to the well known de Almeida–Thouless line $T_f \propto H^{2/3}$ suggesting an acentric long range spin-glass behavior and mean-field system.

BiFeO$_3$ (BFO) is unusual or perhaps unique in that it exhibits magnetism and ferroelectricity at room temperature. Ferroelectromagnetic materials, i.e., multiferroics, exhibit ferroelectric (or antiferro-electric) properties in combination with ferromagnetic (or antiferromagnetic) properties.[1,2] BFO is a rhombohedrally distorted ferroelectric perovskite (T$_c$ ≈ 1100K) with space group R3c,[3,4] which permits coupling between magnetism and ferroelectricity. However, BiFeO$_3$ shows G-type antiferromagnetism up to 643K (T$_N$),[3-5] in which all neighboring magnetic spins are oriented antiparallel to each other, and addition, the axis along which the spins are aligned precesses throughout the crystal, resulting in a modulated spiral spin structure with a long periodicity of ~ 620Å.[4,5] This cycloidal spin modulation is thought to cause linear magnetoelectric coupling to average spatially to zero in single-crystals; however, this modulated spin structure was once thought considered to be absent in constrained films. As a result, weak ferromagnetism has been suggested experimentally[6-9] and also predicted theoretically [10] in thin films. For example, Wang and co-workers[6] fabricated an epitaxially constrained BFO film that exhibited pronounced thickness dependence of ferromagnetism. More recent studies,[8,11] however, show clearly that the compressive epitaxial strain does not enhance the magnetization in BFO films, and Lebeugle *et al.*, for example,[11] note that just 1% mol of paramagnetic Fe$^{3+}$ (probably due to presence of Bi$_{25}$FeO$_{39}$) can account for all the low-temperature magnetic enhancement in their single crystals, and that removing such impurities with HNO$_3$ removes virtually all traces of ferromagnetism in their samples.

A survey of the literature reveals a dramatic change in the magnetic properties of BFO at temperatures below 200 K.[8,9,13-15] Recently we inferred a spin glass transition below 120K, which follows mean field theory[16] and is similar in some respects to those in other orthoferrites. Latter we observed magnons in BFO by inelastic light scattering techniques,[17] showing the spin wave behaviour near the transition temperatures 140K and 201K and the enhancement the Raman intensity of the magnon.[18-20] Cazayous et al.[21] also reported strong magnon anomalies at the same temperatures. Spin reorientation (SR) transitions in orthoferrites have been extensively studied.[22,23] For example, in ErFeO$_3$ Koshizuka and Ushioda[22] observed two one–magnon branches by inelastic scattering technique showing the frequency dependence near the transition temperature and the enhancement the magnon intensity. More recently Redfern et al.[24] noted that transition near 200K shows strong magnetoelastic coupling in the Hz regime whereas anomalies near 140 K show strong elastic coupling in the MHz regime in data from Carpenter et al.[24]

The conventional wisdom (Young et al.[25]) is that short-range Ising-like spin glasses cannot have a de Almeida-Thouless line (AT-line), and that the critical exponent $z\nu = 8.0 \pm 1.0$. However, Fischer and Hertz[26] point out that acentric (non-centrosymmetric) ferroelectric magnetic spin glasses cannot be Ising-like and probably violate other predictions of standard spin-glass theory. The original spin glass model of Karkpatrick and Sherrington[27] described magnetic systems within a mean field theory, for which the critical exponent $z\nu = 2$ characterizes the frequency dependence of a characteristic freezing temperature. Recent work has generally applied Ising statistics to such spin glasses, But $BiFeO_3$ is a unique case of an asymmetric spin glass exhibiting critical exponent $z\nu = 1.4$ similar to the mean-field prediction of 2.0.

In the present work we report magnetic and phonon properties of the pseudo-cubic $[111]_c$-oriented rhombohedral BFO thin films that have been known to possess giant spontaneous polarizations ($P_s$) along the $[111]_c$ axis[28] with the easy magnetization plane perpendicular to this pseudo-cubic $[111]_c$ axis (or equivalently perpendicular to hexagonal $[001]_h$).[29] We observed a splitting in the ZFC and FC magnetization curves at a characteristic temperature, $T_{irr}$. This splitting temperature was strongly dependent on the applied magnetic field, typical of a spin-glass-like transition.

BFO thin films were grown on (111) STO substrates by employing pulsed laser deposition (PLD) method.[30] The average thickness of these films, as estimated using field-emission scanning electron microscopy, was 300±3 nm. To examine the structure of the PLD-grown BFO film on a STO (111) substrate, theta-2-theta ($\theta$-$2\theta$) x-ray diffraction (XRD) and Φ-scan experiments were carried out, and their results are shown in Fig. 1. The pattern reveals purely $[111]_c$-oriented rhombohedral BFO reflections. The degree of in-plane orientation was assessed by examining XRD Φ-scan spectra. As presented in the inset of Fig. 1, the peaks for (022) reflection of the $[111]_c$-oriented domain occur at the same azimuthal Φ angle as those for STO (022) reflection and are 120° apart from each other. This clearly indicates the presence of threefold symmetry along the [111]$_c$ direction and a coherent epitaxial growth of the BFO film with R3c symmetry on a STO (111) substrate. A superconducting quantum-interference-device-based magnetometer (Quantum design MPMS-5) was used for the magnetization measurements which were carried out by cooling the sample to a desired temperature in the presence or absence of an applied magnetic field.

Figure 2 displays the ZFC and FC magnetization curves of the epitaxial BFO film having rhombohedral R3c symmetry. The substrate effect from STO(111) has been substracted from the magnetization data. During the measurements, an external magnetic field of 10 kOe (1$T$) was applied parallel to the out-of-plane [111]$_c$ direction. The magnetization induced along the in-plane direction which is perpendicular to [001]$_h$ (i.e., [111]$_c$) was measured because the magnetization easy axis of R3c BFO is parallel to [110]$_h$ which is vertical to [001]$_h$.[29] As shown in Fig. 2, the ZFC and FC magnetizations gradually increase with decreasing temperature, which is presumably caused by local clustering of spins.[12] The most prominent feature of Fig. 2 is that there is a large discrepancy between the ZFC and FC curves in the film beginning at ~74.7 K, which increases with decreasing temperature. The observed splitting in the ZFC and FC curves at low temperatures is a hallmark of spin-glass-like transition. In addition to this, we have also observed a sharp cusp at around 50 K in the ZFC curve, which can be attributed to a typical blocking process of an assembly of superparamagnetic spin moments.[31] On the contrary, these moments are aligned parallel to the applied field during the FC measurement, leading to a large discrepancy between the FC and ZFC curves below the freezing temperature. Other researchers have very recently reported aging within the inferred spin-glass temperature range. Shvartsman et al.[31] confirm non-ergodic behavior of the low-field magnetization at low T. They suggest that that this might be a reentrant phenomenon, since the system being primarily antiferromagnetic reveals a spin spiral counteracting the formation of weak ferromagnetism due due to global spin canting. However, they exclude a generic spin glass phase, "since only cumulative relaxation is found after isothermal aging below T$_g$ instead of classic hole burning and rejuvenation."

The ZFC and FC magnetization characteristics of the epitaxial BFO film were further examined by applying the bias magnetic field with various strengths. The results are presented in Fig. 3(a) with the field strength of 0.5, 1.0, 1.5, 3.0, 5.0, 7.0 and 10.0 kOe in ascending order. The splitting temperature $T_{irr}(H)$ (irreversibility in ZFC) gradually decreases with increasing field strength. The splitting is accompanied by the observation that the cusp maximum becomes smeared out with decreasing field strength [Fig. 3(a)]. This indicates that the magnetic energy at a high-field becomes sufficient to overcome the energy barrier between possible equilibrium orientations of the magnetic moments, thereby decreasing $T_{irr}(H)$. This observation also supports the spin-glass-like behavior of the present BFO film, which arises from the spin reorientation in the easy magnetization plane. Similar behavior was also reported by Park *et. al.*[14] in their ZFC and FC curves of BFO nanoparticles, suggesting a spin-glass-like transition at low temperatures.

To test the validity of the present spin-glass model of the [111]$_c$-oriented BFO film, we have examined the field-dependent freezing temperature, $T_f(H)$, using the Almeida-Thouless(AT) equation,[12,32] namely, $H = A_{AT}[1-\{T_f(H)/T_f(0)\}]^{3/2}$, where $T_f(H)$ is equal to $T_{irr}(H)$ which corresponds to the onset of the irreversible behavior under the field $H$. The data fitting by the AT equation for higher magnetic fields (1.5~10 kOe) yields $T_f(0)$ = 140 K, as shown in Fig 3(b). Note that this freezing temperature agrees within a small uncertainty with the temperature [18-20] at which the magnon cross section diverges T = 140.3K. From the present data, however, it is very difficult to assign an exact spin-glass temperature in this cross-over regime because sharp transitions at higher fields, i. e. the presence of Almeida–Thouless line stability does not favor the short range Ising-type spin configuration,[25] whereas the Heisenberg-type ferromagnet with non-commuting spin operators is likely to be more appropriate at lower fields.[26] In Ref [16], we reported that the critical exponent describing the slowing down of the glassy dynamics $z\upsilon \approx 1.4$, which is much closer to the value excepted in a mean field system (where $z\upsilon \approx 2.0$) then in the classical short range Ising magnetic spin glass ($z\upsilon \approx 7-10$). La$_{0.5}$Mn$_{0.5}$FeO$_3$ is a good example of such a non- standard spin glass with $z\upsilon \approx 1.0$.[33] Fischer and Hertz[26] have emphasized that no published theories apply to spin glasses lacking an inversion center, and further, that such glasses can not possibly be Ising–like. On this basis it can be suggested that the spin glass transition is coupled with a long-range order parameter (strain) responsible for its mean field behavior and that the symmetry is acentric.

The behavior of magnetization with temperature is in good agreement with the electromagnon description of BFO.[18-20] These authors inferred spin reorientation at 140 and 200K which are temperatures very close to the predicted transition temperature in this work. In Ref [18] we see both sigma (FM-magnon) and gamma (AFM-magnon) modes at 18.3 cm$^{-1}$ and 26.4 cm$^{-1}$(80K) which also reveal the presence of spin glass behavior in BiFeO$_3$. They are very similar to those two branches in orthoferrites such as ErFeO$_3$.[22] However, at exactly 201K there is an abrupt change in the frequency and intensity of the sigma mode and the gamma mode disappears. This suggests a spin reorientation, as is common in orthoferrites. Therefore there appear to be subtle and unpredicted magnetic changes going on in the region of 201K, far below $T_N$ = 640K and these may influence the spin-glass behavior we see initially on cooling at 140K. To rephrase this important point: Spin-glass phases are usually not so far below the Neel temperature in magnetic materials where they occur at all; previously this cast doubt on spin glass behaviour in BiFeO$_3$ at cryogenic temperatures, since $T_N$ > 630K. However, our recent discovery of low-temperature spin-reorientation transitions makes it more plausible.

In conclusion, we examined the ZFC and FC magnetization curves of the [111]$_c$-oriented epitaxial BFO film with R3c symmetry. The two curves showed a discrepancy beginning at a characteristic temperature, $T_f(H)$, which did depend on the applied field, revealing spin-glass-like behavior. However, it is known [34] that such an H$^{2/3}$ dependence is not in itself proof for a spin-glass state, which can also arise from superparamagnetic behaviour. Therefore, other data, including the frequency dependence of the temperature peak in susceptibility and the Vogel-Fulcher dependence,[16] as well as aging phenomena,[31] are helpful in inferring a glassy state. Note also in this context that antiferromagnetic ordering and spin-glass phenomena may coexist both experimentally and in mean field models.[35]

In general it is not easy to prove the existence of a spin glass: An AT-line can occur in superparamagnets; aging and rejuvenation (which we will show for bismuth ferrite in a separate paper) can occur in any ferroic system with domain pinning; and frequency dependent susceptibilities can occur in relaxors. Therefore numerous separate experiments are required, but we believe the AT-line shown here will be very helpful in this regard.

We gratefully acknowledge the financial support from the DOD grants W911NF –06–1–0030 and W911NF-06-1-0183. The magnetic measurements were supported by NoE FAME (FP6-500159-1), the STREP MaCoMuFi (NMP3-CT-2006- 033221) from the European Community and by the CNRS, France.

## * Figure Captions

**FIG. 1.** Theta-2-theta ($\theta$-$2\theta$) XRD pattern of a PLD-grown BiFeO$_3$ thin film on a SrTiO$_3$ (111) substrate with the intensity profile in logarithmic scale. The inset presents Φ-scan diffraction patterns on (022) planes.

**FIG. 2.** Temperature dependence of the dc magnetization (ZFC and FC) of the [111]$_c$-oriented BiFeO$_3$ thin film measured under the applied magnetic field of 10 kOe along the out-of-plane [111]$_c$ direction.

**FIG. 3. (a)** Temperature dependence of the dc magnetization (ZFC and FC) of the [111]$_c$-oriented BiFeO$_3$ thin film measured at various strengths of the applied magnetic field (0.5, 1.0, 1.5, 5.0, and 10.0 kOe in ascending order). **(b)** Experimentally observed values of $T_{irr}$ fitted with the Almeida-Thouless (AT) line equation.

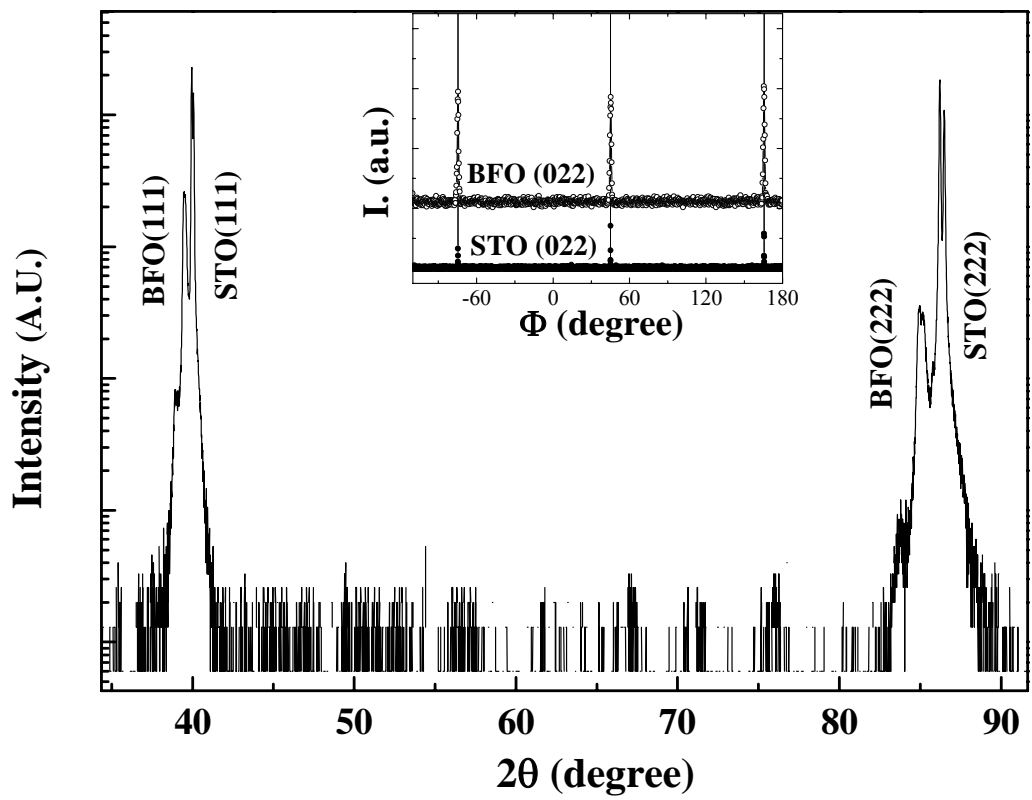

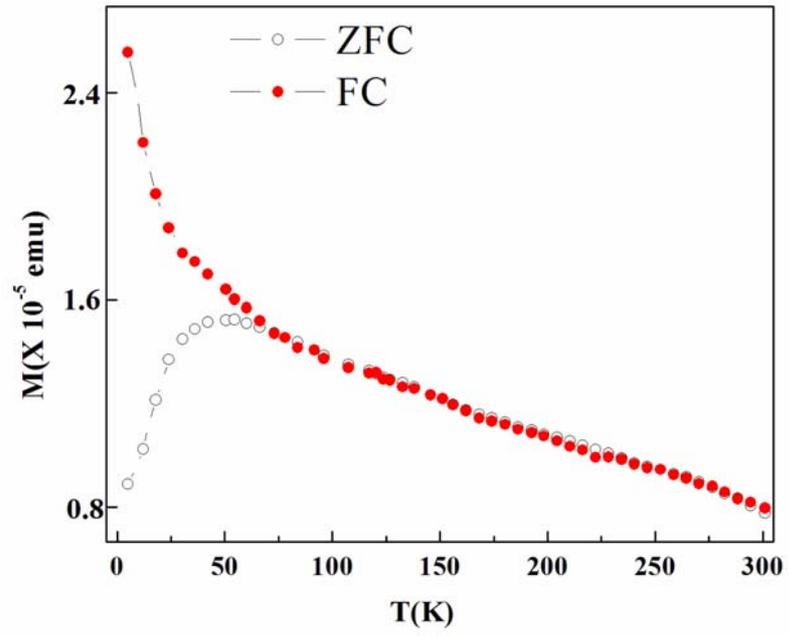

**Figure:2**

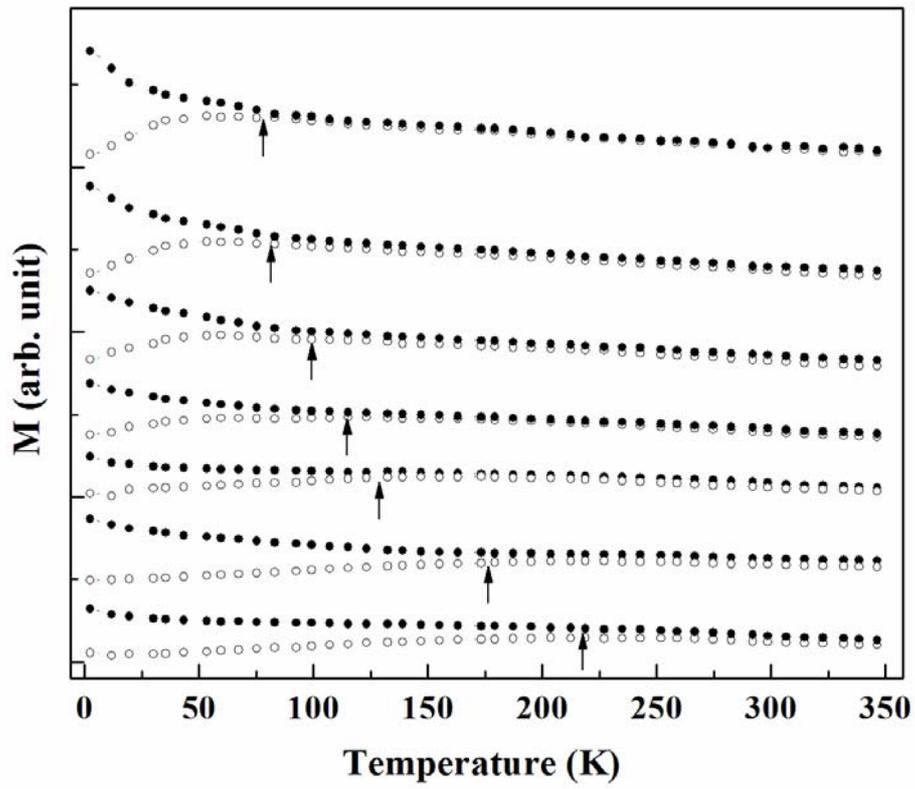

**Figure:3(a)**

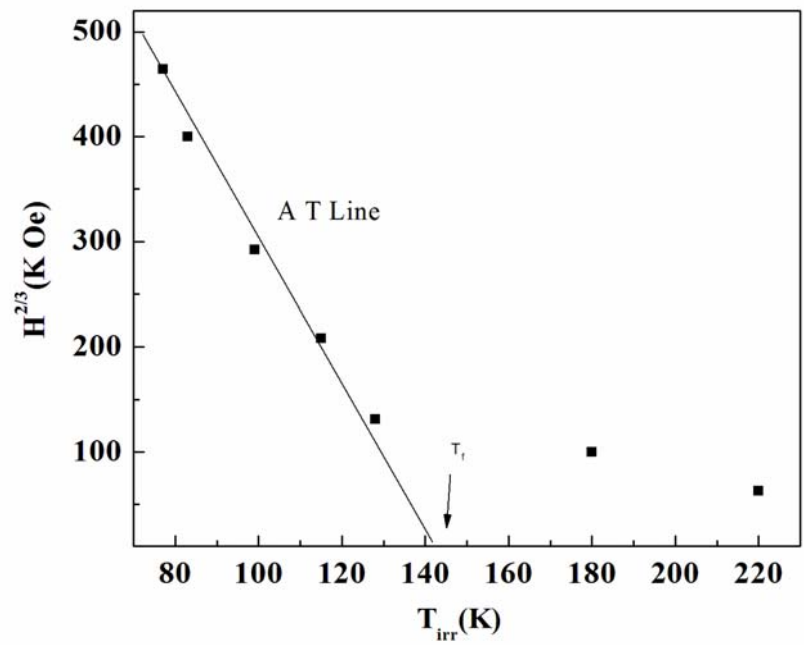

**Figure:3(b)**